\documentclass{aip-cp}

\usepackage[numbers]{natbib}
\usepackage{rotating}
\usepackage{graphicx}

\begin{document}

% Title portion
\title{$\Lambda p$ Elastic Scattering in the CLAS Detector}

\author[aff1]{John W. Price\corref{cor1} for the CLAS Collaboration
\eaddress[url]{https://www.csudh.edu/physics}}
%\author[aff2,aff3]{Author's Name\noteref{note2}}
%\eaddress{anotherauthor@thisaddress.yyy}

\affil[aff1]{California State University, Dominguez Hills, 1000 E. Victoria St., Carson, CA 90747}
%\affil[aff2]{Additional affiliations should be indicated by superscript numbers 2, 3, etc. as shown above.}
%\affil[aff3]{You would list an author's second affiliation here.}
\corresp[cor1]{Corresponding author: jprice@csudh.edu}
%\authornote[note1]{For the CLAS Collaboration.}
%\authornote[note2]{This is an example of second authornote.}

\maketitle

\begin{abstract}
The elastic scattering process $\Lambda p\to\Lambda p$ offers insights
on multiple problems in nuclear physics. $\mathrm{SU}(3)_F$ symmetry
implies a close agreement between the $\Lambda p$ and $pp$ scattering
cross sections. The $\Lambda p$ elastic scattering cross section can
also illuminate the structure of neutron stars. A data-mining project
was started using multiple CLAS data sets taken for other purposes
with photon beams on a long liquid hydrogen target. A $\Lambda$
produced in a process such as $\gamma p\to K^+\Lambda$ can interact
with a second proton inside the target before either decaying or
leaving the target. The good angular acceptance and momentum
resolution of CLAS make it well-suited for this type of analysis, even
though it was not designed for such a measurement. The scattered
$\Lambda$ can be identified from the $\pi^-p$ invariant mass. The 
four-vector of the initial $\Lambda$ is then reconstructed in the
process $Xp\to\Lambda p$, which shows a strong peak at the $\Lambda$
mass with roughly twice the number of events as the existing world
data sample. This observation opens up the possibility of other
measurements using secondary beams of short-lived particles. This
paper will discuss the current status of the analysis, and our plans
for future work on this project.
\end{abstract}

% Head 1
\section{MOTIVATION}
The hyperon-nucleon interaction is of fundamental importance in both
nuclear physics and astrophysics. The so-called ``hyperon puzzle'' in
neutron star structure is based, in part, on experimental measurements
of the hyperon-nucleon cross section, which are badly in need of
improvement.   $\mathrm{SU}(3)_F$ symmetry, a powerful tool in nuclear
physics, treats all hadrons in the same SU(3) multiplet as different
manifestations of the same particle.  This implies a relationship
between the $\Lambda N$ and $NN$ interactions that has only been
poorly tested.  One of the main reasons for this is the short lifetime
of the hyperons.  The longest-lived hyperon, the $\Lambda$, has a
decay length given by $c\tau_{\Lambda}=7.89\,\mathrm{cm}$. 

To improve the state of our understanding of this interaction, we have
developed a new technique, in which the incident $\Lambda$ (referred
to here as the ``beam'' $\Lambda$) is produced via a process such as
$\gamma p\to K^+\Lambda$.  The beam $\Lambda$ can then interact with a
second proton inside the target before either decaying or leaving the
target.  This technique has never been attempted before with a photon
beam.

\subsection{The Hyperon Puzzle}
A neutron star is a gravitationally bound massive object, primarily
consisting of neutrons.  As the mass of such an object increases, the
Pauli pressure of the nucleons (effectively, the chemical potential of
the neutrons in the star) also increases.  At high enough masses,
this pressure can be relieved by the transmutation of neutrons into
hyperons.  This has the effect of softening the neutron star's
equation of state (EoS), reducing the maximum mass a neutron star can
attain before collapsing to a black hole.  Many EoS calculations lead
to a maximum mass that is less than the masses of already-observed
neutron stars.  

The conflict between the theoretical expectation of
the existence of hyperons in neutron stars and the effect that
existence has on the EoS and, in turn, the maximum observed mass, is
known as the \emph{hyperon puzzle}.  There have been many theoretical
attempts to resolve the hyperon puzzle.  These focus primarily on
introducing a repulsive force, often due to the $\Lambda NN$
interaction, to help to stiffen the EoS.\cite{Lon15}  However, more
data are needed to constrain the hyperon-nucleon interaction.

\subsection{$\mathrm{SU}(3)_F$ Symmetry}
The strong nuclear interaction is governed by the QCD Lagrangian
${\cal L}_{QCD}$, which can be rewritten as the sum of two terms
\cite{RPP18}:  
\begin{equation}
  {\cal L}_{QCD} = {\cal L}_0 + {\cal L}_m,
\end{equation}
where ${\cal L}_m$ represents the single term that depends on the
quark mass $\sum_q\overline{\psi}_{q,a}m_q\delta_{ab}\psi_{q,b}$, and
${\cal L}_0$ contains the rest of the Lagrangian.  In the limit of
massless quarks, the QCD Lagrangian reduces to the ${\cal L}_0$ term.
This leads to the prediction that, in this limit, all of the baryons
in a given multiplet will have the same mass, and have similar
properties.  Baryons in different multiplets will have different 
masses from each other, due to the dynamics of the strong interaction.
Within each multiplet, we find that the different masses of the
baryons are determined by the mass term ${\cal L}_m$.

This relatively simple idea works surprisingly well.  We find, for
instance, that the mass splitting between the ground-state and the
first excited-state nucleon, the N(1440), is approximately 500 MeV.
The corresponding mass splitting in the $\Lambda$ sector compares the
ground-state $\Lambda(1115)$ to the first excited-state (octet)
$\Lambda$, the $\Lambda(1600)$, for a splitting of 485 MeV.  The
difference in these mass splittings is only 15 MeV, remarkably close
for strong interaction calculations.

\begin{figure}[t]
  \centerline{\includegraphics[width=250pt]{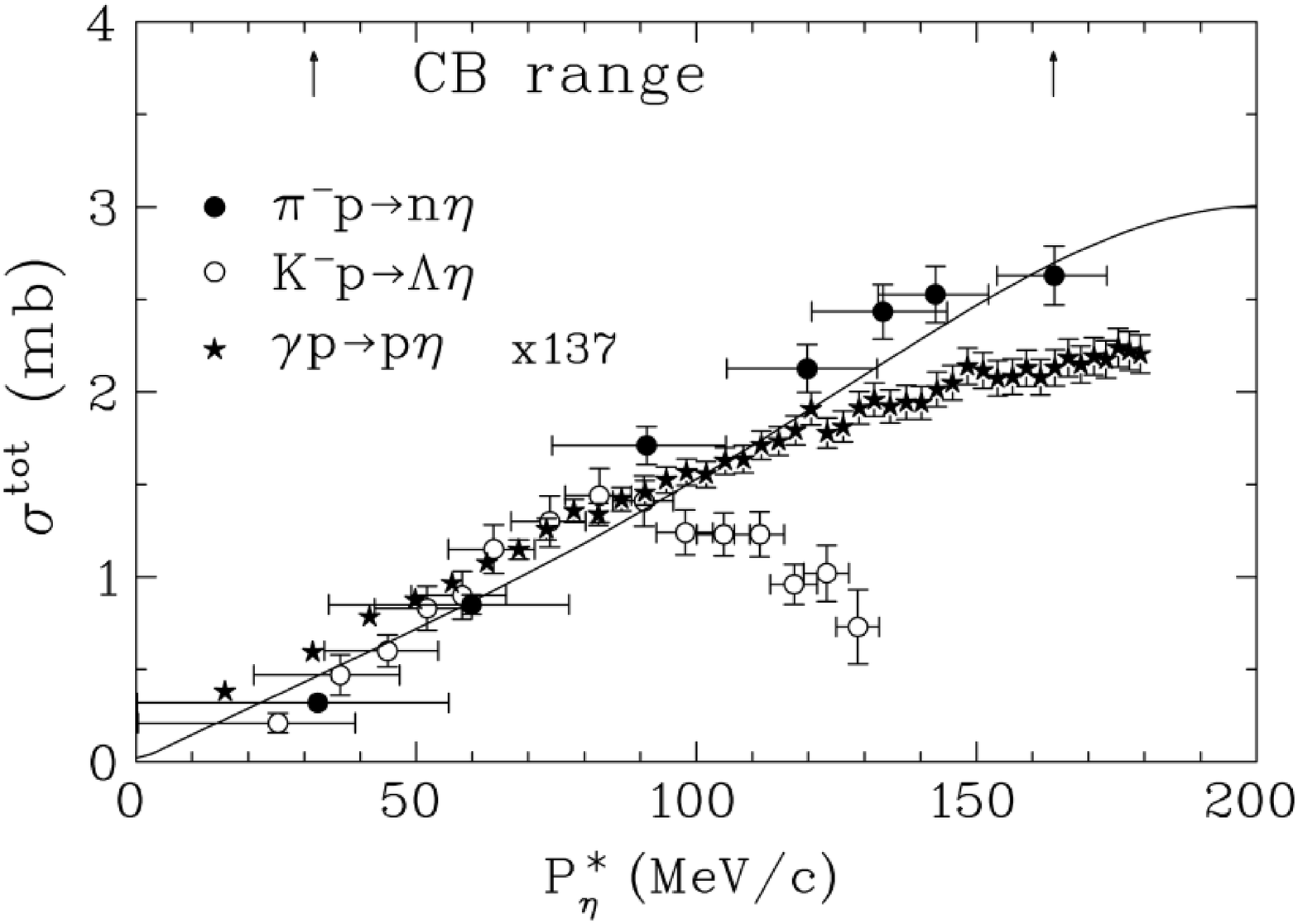}}
  \caption{\label{fig:SU3Symmetry}The $p^*_{\eta}$ dependence of
    $\sigma^{tot}(\pi^-p\to\eta n)$,
    $\sigma^{tot}(K^-p\to\eta\Lambda$), and $\sigma^{tot}(\gamma
    p\to\eta p$).  Experimental data for $\pi^-p\to n\eta$ (filled
    circles) are from~\cite{Pra05}; data for $K^-p\to\eta\Lambda$
    (open circles) are from~\cite{Sta01}, and data for $\gamma
    p\to\eta p$ (stars) are from~\cite{Kru95}. The photoproduction
    data have been multiplied by a factor of 137. The solid line shows
    the FA02 prediction of Ref.~\cite{Arn04}.} 
\end{figure}

A similar comparison can be made for cross sections of processes
related by $\mathrm{SU}(3)_F$ symmetry.  Figure~\ref{fig:SU3Symmetry}
shows the cross section for the processes $\pi^-p\to n\eta$ and
$K^-p\to\Lambda\eta$.  
As seen in the Figure, the cross sections for the two processes
$\pi^-p\to n\eta$ and $K^-p\to\Lambda\eta$ are nearly identical near
the threshold, when plotted as a function of the center-of-mass
momentum of the $\eta$.

This leads to the prediction that the cross sections for other
processes related by $\mathrm{SU}(3)_F$ symmetry should be similarly
related.  This can be studied, for instance, in the processes $pp\to
pp$ and $\Lambda p\to\Lambda p$.  A hint of the relationship between
these processes is evident in the Additive Quark Model of Levin and
Frankfurt~\cite{Lev65}.  Using this model, one obtains the following
relationship:
\begin{equation}
  \label{eq:AdditiveQuarkModel}
  \sigma_{\Lambda p} = \frac{1}{2}\left(\sigma_{pp}+\sigma_{\Xi p}\right)
\end{equation}
While this model is best-suited to high-energy measurements, it is
still a good place to start.

\subsection{Previous Measurements}
\begin{figure}[t]
  \centerline{\includegraphics[width=0.95\textwidth]{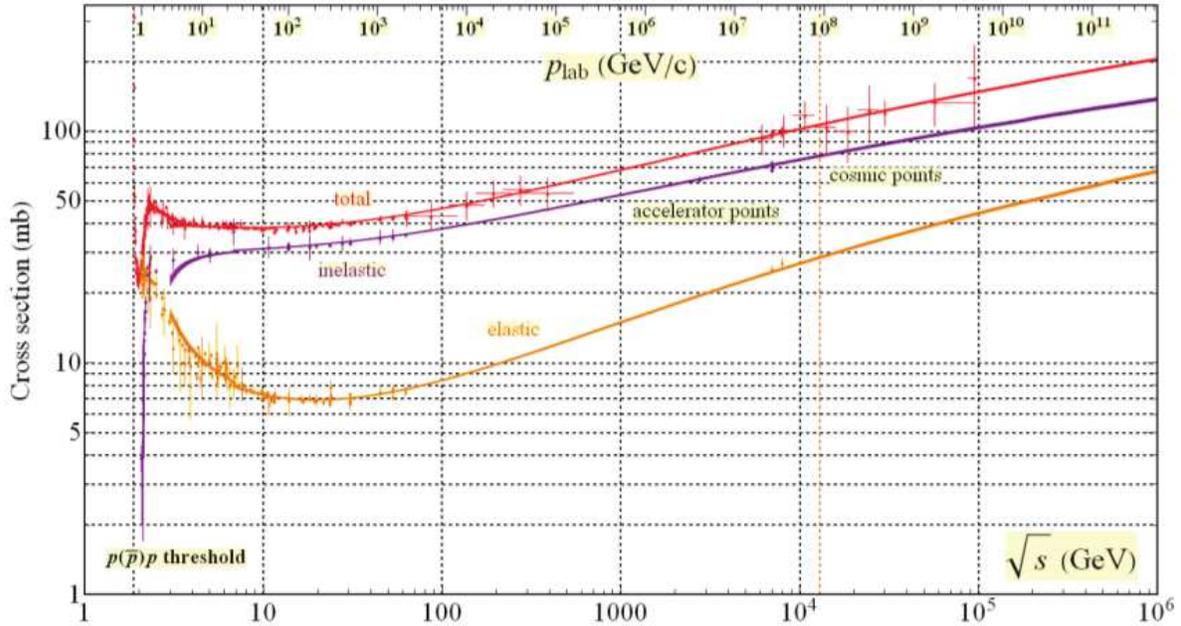}}
  \caption{\label{fig:PreviousDatapp}The existing data for the process
    $pp\to pp$, from Ref.~\cite{RPP18}.}
\end{figure}

\begin{figure}[t]
  \centerline{\includegraphics[width=0.95\textwidth]{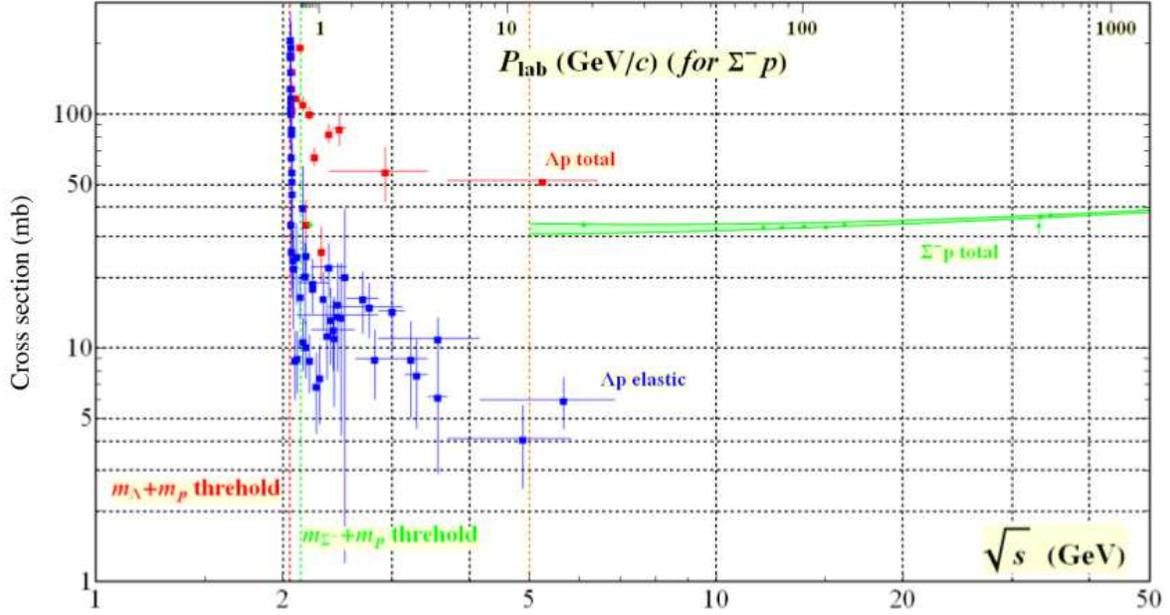}}
  \caption{\label{fig:PreviousDataLp}The existing data for the process
    $\Lambda p\to\Lambda p$ (bottom), from Ref.~\cite{RPP18}.}
\end{figure}

The present state of the data for the process $\Lambda p\to\Lambda p$
is poor.  Figure~\ref{fig:PreviousDatapp} shows the state of cross section
measurements for the processes $pp\to pp$;
Fig.~\ref{fig:PreviousDataLp} shows the equivalent cross section data
for $\Lambda p\to\Lambda p$.
As seen in the figures, the quality of the data for the process $pp\to pp$
is of much higher quality than that for $\Lambda p\to\Lambda p$.  

The present world data sample for the process $\Lambda p\to\Lambda p$
consists of thirteen
publications.\cite{Cra59,Ale61,Gro63,Bei64,Pie64,Sec64,Vis66,Bas67,Ale68,Sec68,Kad71,And75,Mou75}
Table~\ref{tab:PreviousData} summarizes the existing world data set
for this process.
\begin{table}[h]
\caption{The existing data for the process $\Lambda p\to\Lambda p$.}
\label{tab:PreviousData}
\tabcolsep7pt\begin{tabular}{lcccc}
\hline
Reference
   & $\Lambda$ source & Detector & $p_{\Lambda}$ & $N_{\Lambda p\to\Lambda p}$ \\
\hline
Crawford \emph{et al.}~\cite{Cra59}
   & $\pi^-p\to\Lambda K^0$ & LH$_2$ BC & 0.5--1.0 & 4 \\
Alexander \emph{et al.} (1961)~\cite{Ale61}
   & $\pi^-p\to\Lambda K^0$ & LH$_2$ BC & 0.4--1.0 & 14 \\
Groves~\cite{Gro63}
   & $K^-N\to\Lambda\pi$ & Propane BC & 0.3--1.5 & 26 \\
Beilli\`ere \emph{et al.}~\cite{Bei64}
   & $K^-N\to\Lambda\pi$ & Freon BC & 0.5--1.2 & 86 \\
Piekenbrock and Oppenheimer~\cite{Pie64}
   & $K^-A\to\Lambda X$ & Heavy Liquid BC & 0.15--0.4 & 11 \\
Sechi-Zorn \emph{et al.} (1964)~\cite{Sec64}
   & $K^-p\to\Lambda X$ & LH$_2$ BC & 0.12--0.4 & 75 \\
Vishnevksi\u{i} \emph{et al.}~\cite{Vis66}
   & $nA\to\Lambda X$ & Propane BC & 0.9--4.7 & 12 \\
Bassano \emph{et al.}~\cite{Bas67}
   & $K^-p\to\Lambda X$ & LH$_2$ BC & 1.0--5.0 & 68 \\
Alexander \emph{et al.} (1968)~\cite{Ale68}
   & $K^-p\to\Lambda X$ & LH$_2$ BC & 0.1--0.3 & 378 \\
Sechi-Zorn \emph{et al.} (1968)~\cite{Sec68}
   & $K^-p\to\Lambda X$ & LH$_2$ BC & 0.1--0.3 & 224 \\
Kadyk \emph{et al.}~\cite{Kad71}
   & $K^-\mathrm{Pt}\to\Lambda X$ & LH$_2$ BC & 0.3--1.5 & 175 \\
Anderson \emph{et al.}~\cite{And75}
   & $p\mathrm{Pt}\to\Lambda X$ & LH$_2$ BC & 1.0--17.0 & 109 \\
Mount \emph{et al.}~\cite{Mou75}
   & $p\mathrm{Cu}\to\Lambda X$ & LH$_2$ BC & 0.5--24.0 & 71 \\
\hline
\end{tabular}
\end{table}
There have been a total of less than 1300 observed $\Lambda
p\to\Lambda p$ events.  All of the experiments used bubble chambers,
which limited the rate at which data could be taken.  In
all of the previous measurements, the incident $\Lambda$ is created
inside a bubble chamber via some other process (the ``$\Lambda$
source'' column in Table~\ref{tab:PreviousData}; the $\Lambda$ then
interacts with a proton within the bubble chamber to produce the
$\Lambda p\to\Lambda p$ event.

\section{DATA-MINING PLAN}
A similar approach to this process could be successful today.  While
the detectors in common usage today do not allow the complete
visualization of events afforded by the bubble chambers used in the
older experiments, they have the advantage of significantly higher rates.

In comparison with the previous data, $\Lambda$s are produced in far
more plentiful quantities in modern experiments using liquid hydrogen
targets.  Because the $\Lambda$ lives for a short time before
decaying, it can interact with a second proton inside the same target.
With a $4\pi$ detector, all the final-state particles can be observed,
allowing the reconstruction of the entire event.

The CLAS detector \cite{Mec03}, at the Thomas Jefferson National
Accelerator Facility (JLab) in Newport News, VA, was originally built to
study the structure of the proton and its excited states.  It is
well-suited for this task.  It has good angular and momentum
resolution, as well as good angular and momentum acceptance.  It also
has a good efficiency for multiparticle final states.  Even though it
was never conceived for such a process, it is difficult to
imagine a detector that could do a much better job.

\begin{figure}[t]
  \centerline{
    \includegraphics[width=0.45\textwidth]{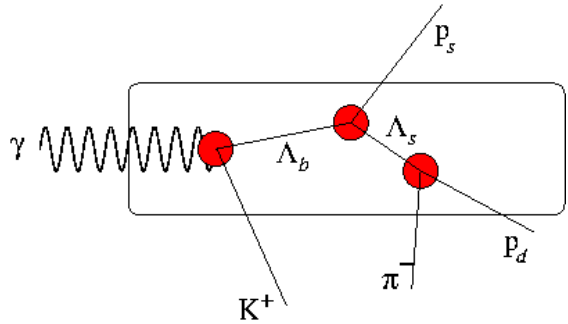}}
  \caption{\label{fig:EventTopology}The event topology for the process
    $\gamma p\to K^+\Lambda;\,\Lambda p\to\Lambda p$ considered in
    this work.}
\end{figure}

In its original layout, the JLab accelerator produced an electron
beam of energies up to 6 GeV in CLAS.\footnote{The accelerator has
  subsequently been upgraded to 12 GeV (11 GeV for CLAS).}  This
electron beam was used to produce a photon beam with a
bremsstrahlung tagging system.  The process that produced the beam
$\Lambda$ was not identified; the simplest such process is $\gamma
p\to K^+\Lambda$.  The complete event topology assuming this
production mechanism is shown in Fig.~\ref{fig:EventTopology}.

With this in mind, a survey of the available data taken by the CLAS
Collaboration was made, to determine a likely dataset in which this
event could be observed.  To decide which dataset to begin this study
with, we looked for one with a large integrated luminosity, with a
long liquid hydrogen target to facilitate the rescattering process.

The CLAS g12 run, taken in the Spring of 2008, was chosen for this
work.  It used a photon beam with energies up to $5.4\,\mathrm{GeV}$,
and took a total of approximately $52\,pb^{-1}$ of $\gamma p$ data.
It had a 40-cm-long liquid hydrogen target, which gives a large number
of potential secondary scattering targets.  The cross section for the
basic production process $\gamma p\to K^+\Lambda$ is approximately
$0.5\,\mu b$; there were thus approximately $2.6\times10^7$ beam
$\Lambda$s for this analysis.

For the process of interest,
\[
\gamma p\to K^+\Lambda;\ \Lambda p\to\Lambda p
\]
the final state was $K^+\pi^-pp$.  The presence of two
protons in the final state, an ``apparent'' violation of baryon
conservation, resulted in an extremely restrictive cut, which
essentially required that some form of rescattering has taken place.

By skimming the data for events with two protons, the size of the
dataset was reduced from 126 TB to approximately 4 TB.  This reduced
dataset could then be analyzed in its entirety very quickly.

\section{EVENT SELECTION}
\begin{figure}[t]
  \centerline{
    \includegraphics[width=0.49\textwidth]{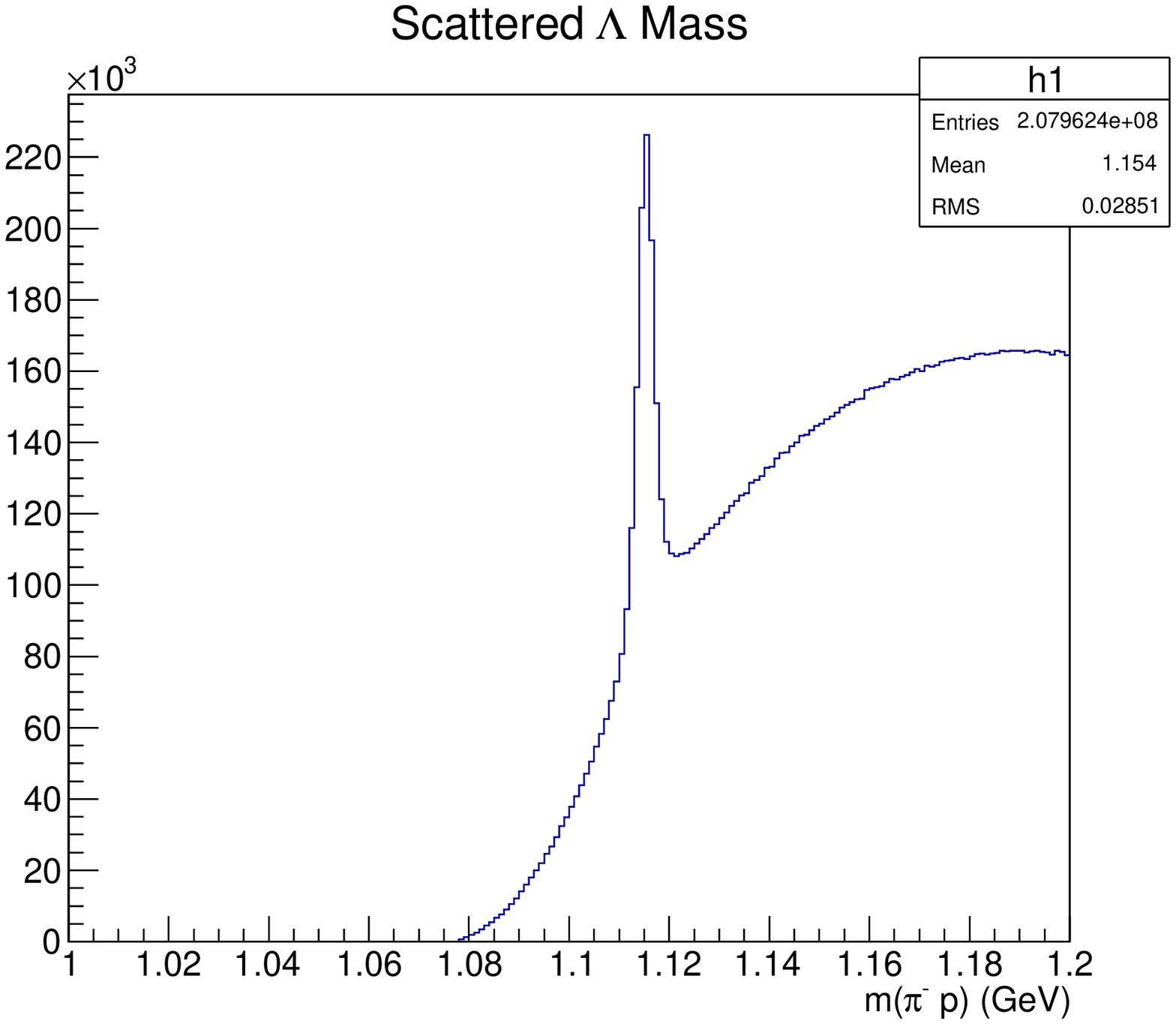}
    \hfil
    \includegraphics[width=0.49\textwidth]{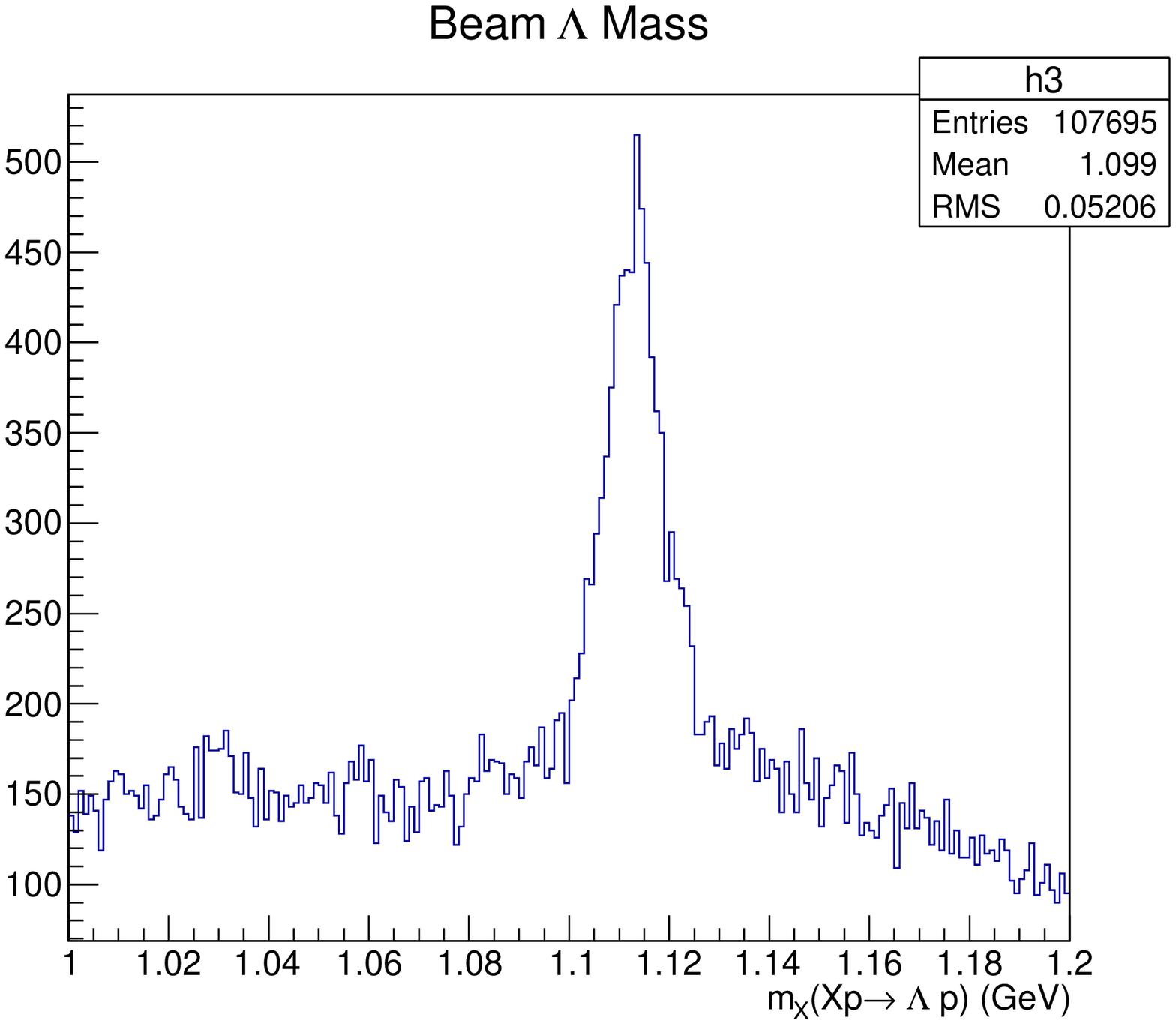}}
  \caption{\label{fig:LambdaMassPlots}Analysis spectra for $\Lambda p$
    elastic scattering.  One-half of the CLAS g12 dataset
    (approximately 25 $\mathrm{pb}^{-1}$ was used in this analysis.
    No attempt has yet been made to reduce the background.  (left) The
    invariant mass $m_{X}$ of the $\pi^-p$ system.  Only the invariant
    mass closest to $m_{\Lambda}=1.115\,\mathrm{GeV}$ is plotted.
    (right) The missing mass $m_{X}$ for the process $Xp\to\Lambda p$.} 
\end{figure}

The data analysis consisted of reconstructing the event from the final
state.  We began by looking for the scattered $\Lambda$
(``$\Lambda_s$'') in the $\pi^-p$ invariant mass.  Because there were
two protons in the final state, there were two such invariant masses to
be checked.  For simplicity, we took the invariant mass closest to
$m_{\Lambda}=1.115\,\mathrm{GeV}$ to indicate which was the ``correct''
proton.  This spectrum is shown in the left plot of
Fig.~\ref{fig:LambdaMassPlots}.
The scattered $\Lambda$ shows up as an unmistakable peak above a large
background which we have not yet attempted to reduce.

If the $\Lambda$ found in this procedure was actually the scattered
$\Lambda$ in the process $\Lambda p\to\Lambda p$, the second proton
provided all the remaining information needed to completely
reconstruct the event.  The beam $\Lambda$ was then the missing
particle in the process $Xp\to\Lambda p$, and should show up in the
corresponding missing mass plot.  This is shown in the right plot of
Fig.~\ref{fig:LambdaMassPlots}.  While we have not done a detailed 
analysis of this plot, it is clear that there are more than twice the
number of events in the peak as are in the entire world data sample.

\section{CROSS SECTION CALCULATION}
The determination of the cross section for this process is more
complicated than for a typical process in nuclear physics.  The
equation for the cross section is given by
\[
\sigma = \frac{N_e}{N_bN_tA\eta} = \frac{N_e}{{\cal L}A\eta},
\]
where $N_e$ is the number of detected events; $N_b$ and $N_t$ are the
number of beam and target particles, respectively; $A$ is the detector
acceptance; and $\eta$ is a catch-all factor for any analysis
inefficiencies.

The product $N_bN_t$ is often referred to as the luminosity
${\cal L}$, and is a measure of the total amount of data available.
For a 
typical experiment, $N_b$ is determined by various beam diagnostic
equipment, either by the accelerator or, in the case of tagged
secondary beams, by the beam tagger system diagnostics.  The number of
target particles $N_t$ is typically calculated using the measured
length of the target, multiplied by appropriate factors of the target
density, atomic mass number, and Avogadro's number
$N_A=6.02\times10^{23}$.

We cannot use such simple diagnostic for $N_b$, however.  There is no
``beam diagnostic'' that can count the number of $\Lambda$s produced
in our detector.  Since the $\Lambda$ is neutral, and decays quickly,
it cannot be measured directly.  While it is possible to ``tag'' the
$\Lambda$ in specific processes, such as $\gamma p\to K^+\Lambda$,
there are many other processes that can produce $\Lambda$s in our
target.

We also cannot use the target length to determine $N_t$, for three
reasons.  First, the beam $\Lambda$ is created within the target; any
protons upstream of the $\Lambda$ production vertex are not available
for later scattering.  Second, the $\Lambda$ will decay after a short
time; any protons downstream of the $\Lambda$ decay vertex are also
unavailable for later scattering.  Third, unlike the beam produced by
the accelerator, the beam $\Lambda$ is not traveling parallel to the
target axis.  It can leave the target through the cylindrical wall
long before traversing the entire target length.

\begin{figure}[t]
  \centerline{
    \includegraphics[width=0.75\textwidth]{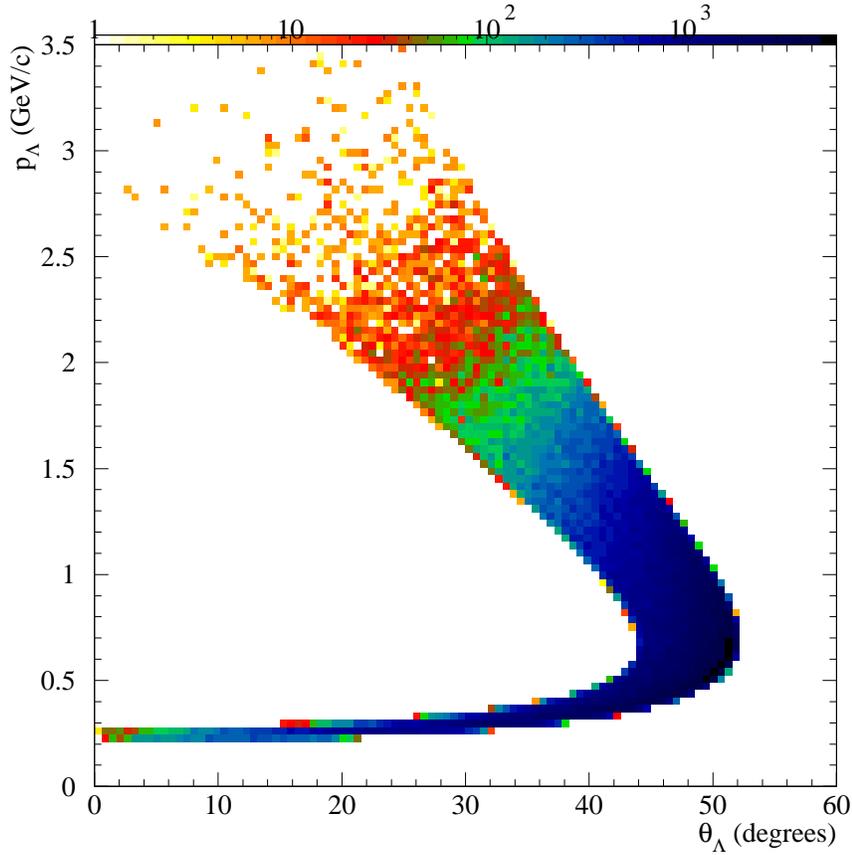}}
  \caption{\label{fig:BeamLambdaKinematics}The kinematics of the
    outgoing $\Lambda$ in the process $\gamma p\to K^+\Lambda$, from
    simulation.  The photon beam energy in the simulation was chosen
    to match the properties of the beam used in the CLAS g12 run.} 
\end{figure}

The best that can be done with respect to $N_b$ is an estimate of the
number of beam $\Lambda$s.  For our first attempt at this, we will
focus on the $\Lambda$ photoproduction process $\gamma p\to
K^+\Lambda$.  Figure~\ref{fig:BeamLambdaKinematics} shows the
kinematics of the $\Lambda$ in the process $\gamma p\to K^+\Lambda$.
This plot is based on the thrown values for the simulation.  For this
purpose, it is not necessary to actually perform the simulation; the
entire calculation is based on the geometry of the target and the
knowledge of the decay length of the $\Lambda$.

To accomplish this, it is necessary to rebin the data.  We are
interested in the flux of $\Lambda$s as a function of the $\Lambda$
momentum $p_{\Lambda}$ and the $\Lambda$ lab angle
$\theta_{\Lambda}$.  In the process $\gamma p\to K^+\Lambda$, however,
the results are normally reported in terms of the photon energy
$E_{\gamma}$ and the kaon c.m.\ angle $\theta^*_{K^+}$.  The
kinematics are overdetermined, so there is a straightforward
relationship that will allow the rebinning of the data.  Once this is
complete, we can make an estimate of how many beam $\Lambda$s are in
each of our kinematic bins, based upon previous measurements of the
cross section for $\gamma p\to K^+\Lambda$.

\begin{figure}[p]
  \centering
  \begin{tabular}{cc}
    \includegraphics[width=0.49\textwidth]{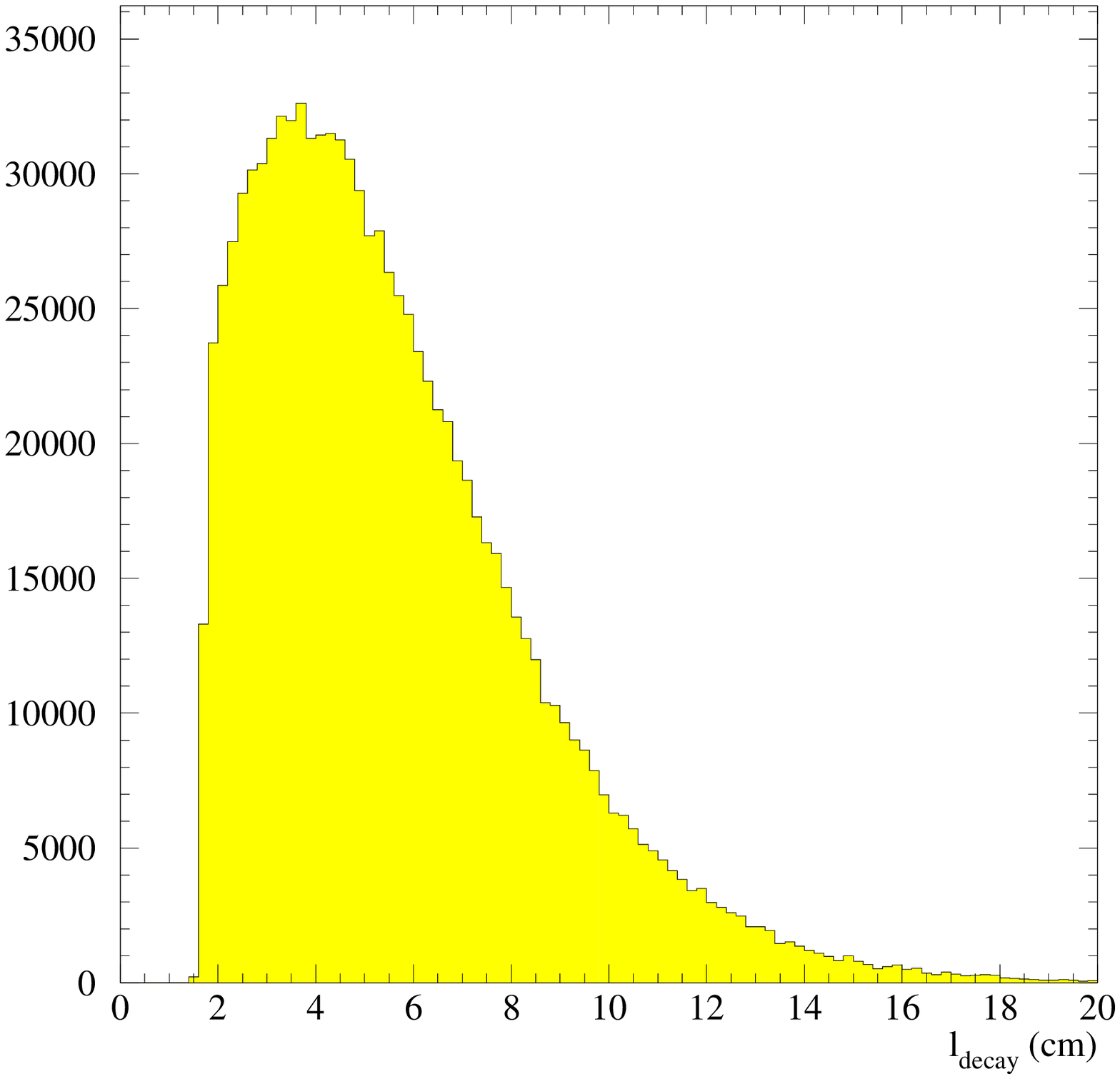} &
    \includegraphics[width=0.49\textwidth]{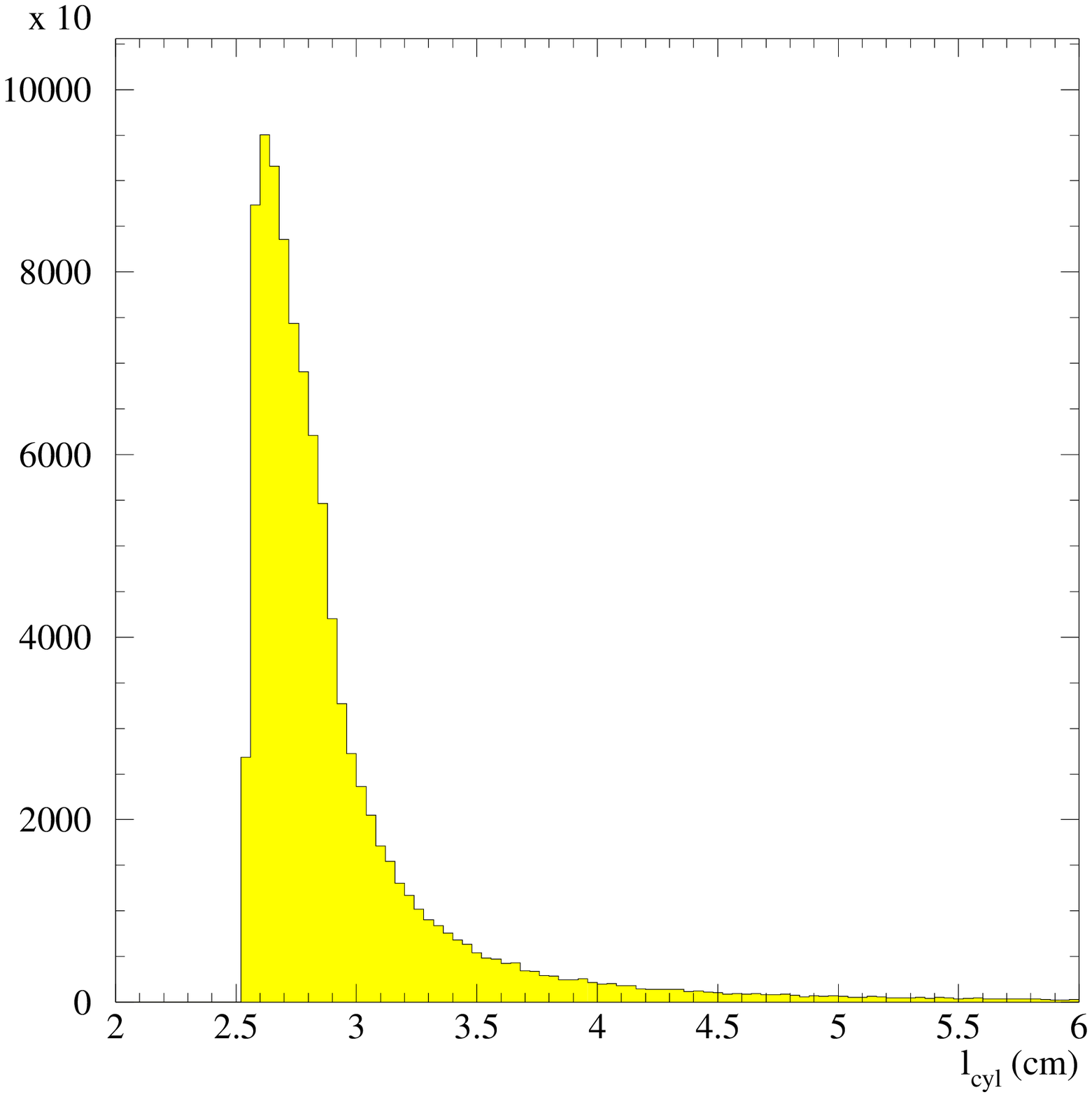} \\
    \includegraphics[width=0.49\textwidth]{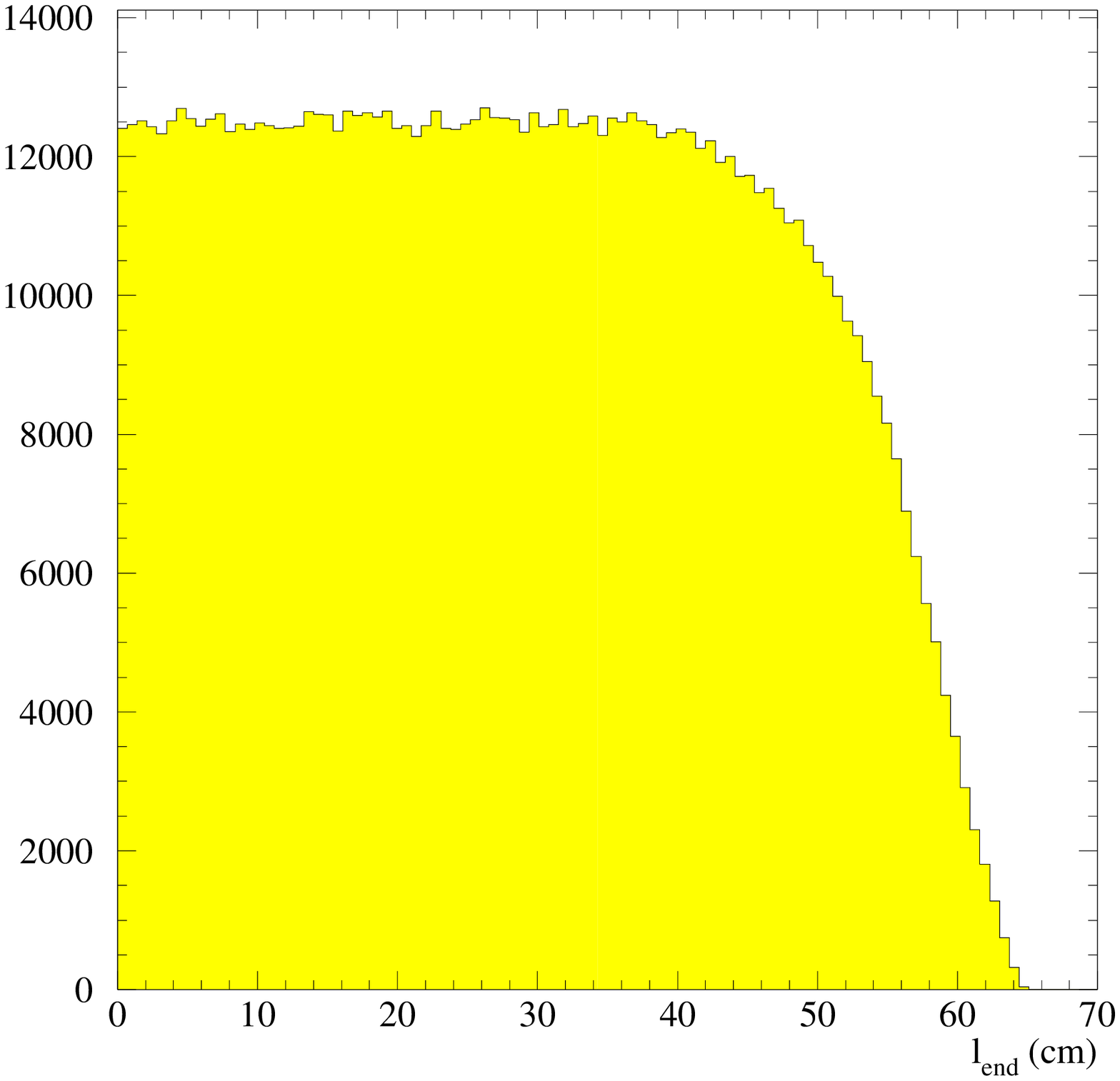} &
    \includegraphics[width=0.49\textwidth]{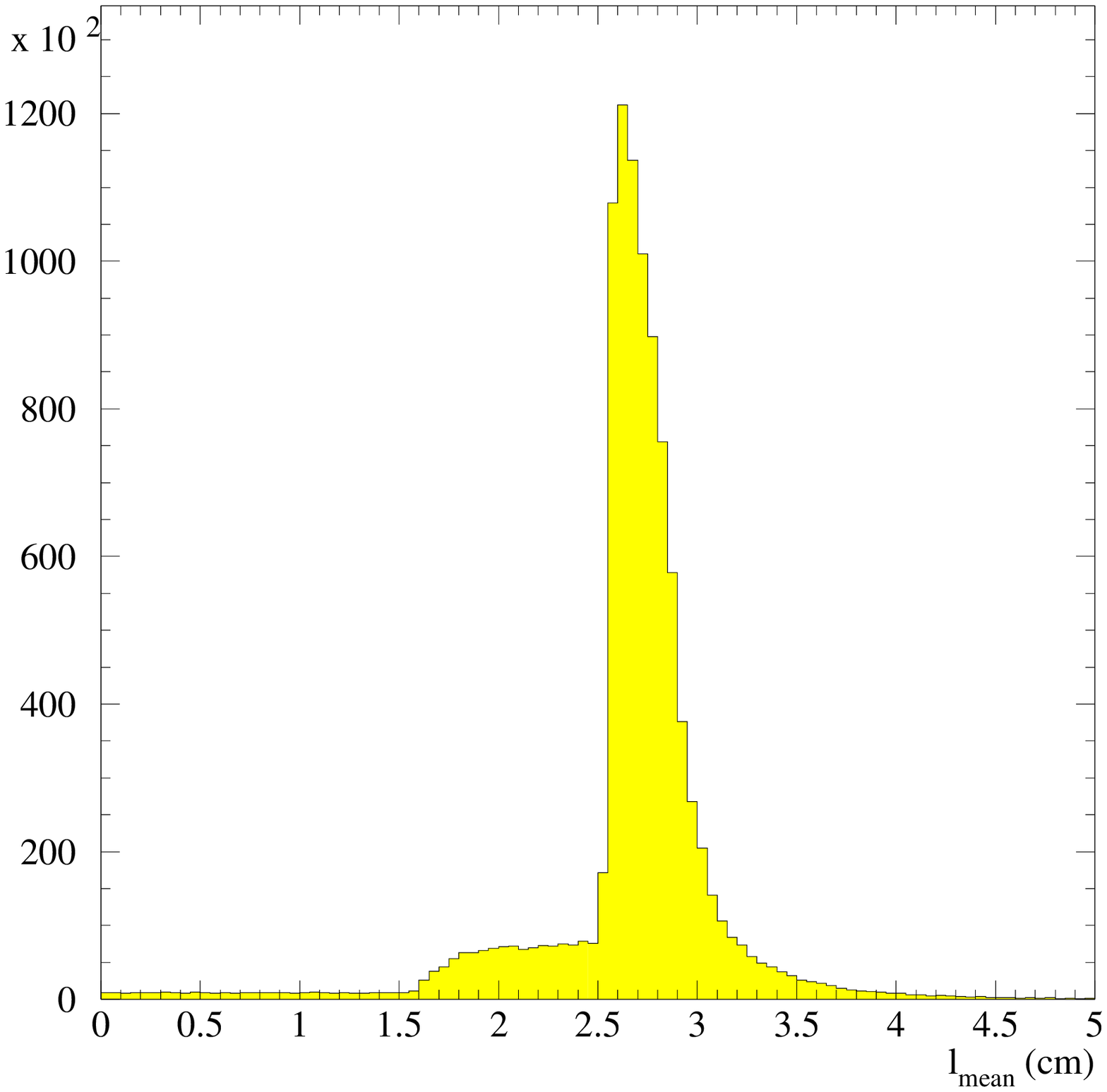}
  \end{tabular}
  \caption{\label{fig:MeanPathLength}The mean path length of the
    beam $\Lambda$ in the g12 liquid hydrogen target.  All values are
    taken from the kinematics of the $\gamma p\to K^+\Lambda$ process
    at the photon energies used in the g12 run.  (top left) The mean
    distance traveled by a $\Lambda$ before decaying.  (top right) The
    mean length traveled by a $\Lambda$ before escaping via the
    cylindrical wall of the target.  (bottom left) The mean length
    traveled by a $\Lambda$ before escaping via the target endcap.
    (bottom right)  The mean path length of all $\Lambda$s in this
    measurement, calculated as the minimum of the other three values.}
\end{figure}

The next step is to determine the effective target thickness.  This
must be done for each of the kinematic bins for our beam $\Lambda$;
two $\Lambda$s with the same momentum will have different mean path
lengths if they are traveling at different angles, since one will
escape the target before the other, for instance.  There are three
possible reasons why the beam $\Lambda$ will not interact with a
second proton in the target: 
\begin{enumerate}
\item The beam $\Lambda$ decays before interacting.  The mean path
  length of a $\Lambda$ of velocity $\beta$ before decaying is given
  by $\ell_{decay}=\beta\gamma c\tau$.  This calculation ignores the
  possibility that the beam $\Lambda$ will leave the target before
  interacting with a proton, effectively assuming a target of infinite
  size.
\item The beam $\Lambda$ is produced at a large angle, and escapes the
  target through the cylindrical wall.  For a particle produced with
  an angle $\theta_{\Lambda}$ relative to the target axis, the mean
  length from the production vertex to the cylindrical wall is given
  by $\ell_{cyl}=r_{tgt}/\sin{\theta_{\Lambda}}$.  This calculation
  ignores the possibility that the beam $\Lambda$ will either 
  decay or exit through the endcap, effectively assuming a target of
  infinite length.
\item The beam $\Lambda$ is produced at a small angle close to the
  downstream end of the target, and escapes through the endcap
  before decaying.  For a particle produced a distance $z_{\Lambda}$
  downstream of the entrance of a target of length $z_{tgt}$ with an
  angle $\theta_{\Lambda}$ relative to the target axis, the mean
  length from the production vertex to the $z$-position of the
  downstream endcap position of the target is given by
  $\ell_{end}=(\ell_{tgt}-z_{\Lambda})/\cos{\theta_{\Lambda}}$.  This
  calculation ignores the possibility that the beam $\Lambda$ will
  either decay or exit through the cylindrical wall, effectively
  assuming a target of infinite radius.
\end{enumerate}
The actual mean path for any given $\Lambda$ $\ell_{mean}$, taking
into account the possibility that it will either decay or exit the
target before interacting with a proton is the minimum of these three
numbers: $\ell_{mean}=min(\ell_{decay},\ell_{cyl},\ell_{end})$.  For
the kinematics of the beam $\Lambda$ in the g12 dataset,
Fig.~\ref{fig:MeanPathLength} shows the distribution of all four
variables $\ell_{decay}$ (top left), $\ell_{cyl}$ (top right),
$\ell_{end}$ (bottom left), and $\ell_{mean}$ (bottom right). 
What we see is that the majority of our $\Lambda$s are escaping
through the cylindrical wall of the target.  Using a long target does
not, therefore, increase the mean path length of the $\Lambda$
noticeably (although it will increase the number of $\Lambda$s
produced). 

Having found both the number of beam $\Lambda$s $N_b$ and the
effective target thickness $N_t$ as a function of the $\Lambda$
momentum $p_{\Lambda}$ and lab angle $\theta_{\Lambda}$, we then can
calculate the luminosity for this measurement using the following
equation:
\[
{\cal L} =
\sum_{all\ bins}N_b(p_{\Lambda},\theta_{\Lambda})N_t(p_{\Lambda},\theta_{\Lambda})
\]

The next step in the calculation of the cross section for $\Lambda
p\to\Lambda p$ is to simulate the acceptance for this process, using
the measured kinematics as input.  This procedure is underway, with
the expectation that it should be complete within the year.

\section{FUTURE DIRECTIONS}
Our success in this endeavor has motivated us to consider other ways
in which this technique can be exploited with detectors such as CLAS.
While we are still in the process of developing a new proposal to take
advantage of the new capabilities of the CLAS12 detector with the
higher energy beam now available at JLab, we are also looking into the
existing data currently stored.  This represents nearly twenty years
of data acquisition, much of which is compatible with studies such as
this.

Our experience with this measurement makes it clear that a long target
is not necessary for such an experiment.  Because the beam particles
nearly always escape the target through the cylindrical walls, the
target length only contributes to the measurement through the number
of beam particles produced.  We are therefore free to look at all
previous datasets for this work.

In principle, any particle produced in sufficient quantities may be
used for such a study.  To begin with, we will focus on processes that
include two protons in the final state.  This will allow us to make
use of the skimmed data already in hand for the $\Lambda p\to\Lambda
p$ study.  Such a skim reduces the size of the background
considerably; from an original size of 230TB, the final skimmed
dataset is slightly more than 4TB.

\subsection{Measurement of $\gamma p\to\Lambda X$}
At the present time, we must restrict ourselves to events in which we
can identify the $K^+$ from the process $\gamma p\to K^+\Lambda$.  The
reason for this is that this process is well-studied, and its cross
section is well-known.  However, it is a relatively small part of the
total $\Lambda$ production cross section at higher energies.  In order
to take advantage of all of the $\Lambda$s produced in CLAS, it will
be necessary to measure the process $\gamma p\to\Lambda X$.  This is
one of our first plans after the present analysis is complete, which
will increase greatly the number of beam $\Lambda$s at our disposal.

\subsection{Confirmation of the $\Lambda p$ scattering result}
Our first effort was to duplicate the measurements made with the g12
dataset.  For this, we look at the g11 dataset.  This set was taken in
2004, and used the same 40-cm-long $\mathrm{LH}_2$ target as the g12
dataset.  The photon energy range was slightly lower, and the trigger
was different, which will affect the events seen by CLAS.  

\begin{figure}[t]
  \centerline{
    \includegraphics[width=0.49\textwidth]{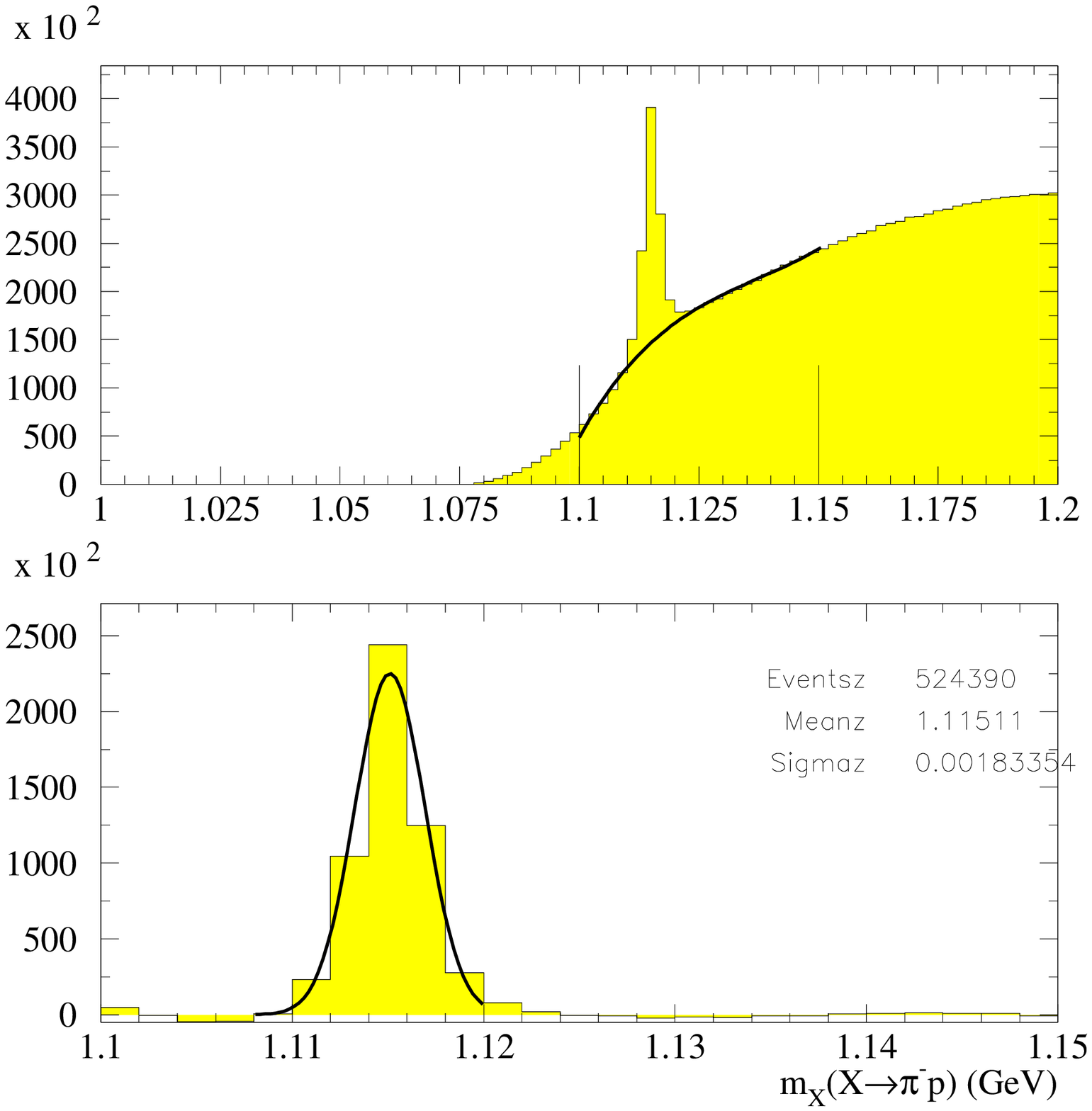}
    \hfil
    \includegraphics[width=0.49\textwidth]{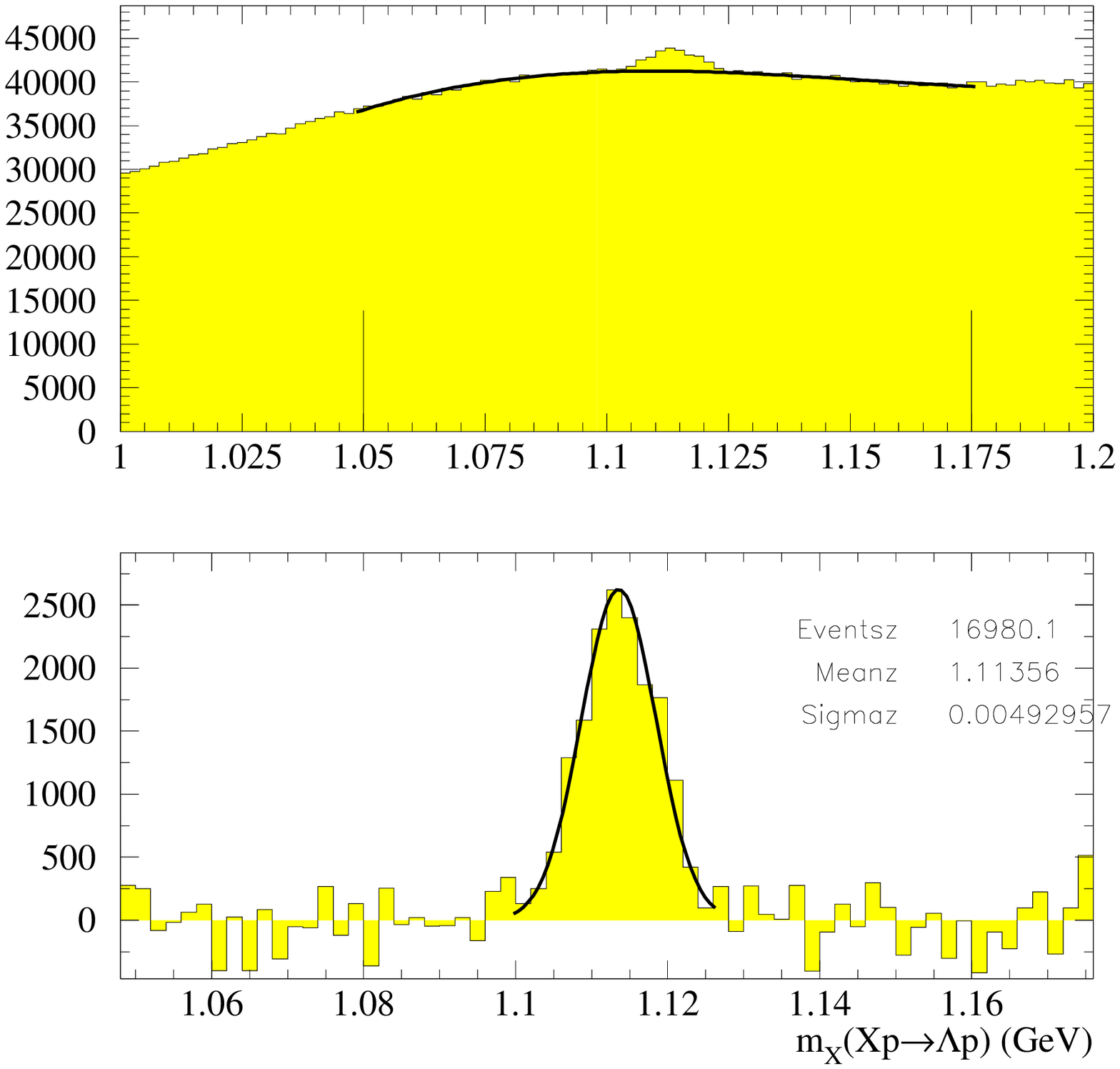}}
  \caption{\label{fig:g11Results}Analysis spectra for the
    analysis of $\Lambda p$ elastic scattering, from the g11 dataset.
    No attempt has yet been made to reduce the background.  The units
    for all plots are GeV.  (left plots)
    The invariant mass $m_{\pi^-p}$ of the $\pi^-p$ system.  Only the
    invariant mass closest to $m_{\Lambda}=1.115\,\mathrm{GeV}$ is
    plotted.  The scattered $\Lambda$ shows up as a strong peak above
    the background.  (right plots) The missing mass $m_X$ for the process
    $Xp\to\Lambda p$.  Only events with $m_{\Lambda_{s}}$ within 5~MeV
    of the known value of $m_{\Lambda}=1.115\,\mathrm{GeV}$ are used.
    (top plots) The raw invariant/missing mass spectra.  The dark line
    is a 3rd-order polynomial fit to the background.  (bottom plots)
    The same plot, with the background subtracted.  The dark line is a
    gaussian fit to the remaining peak.}
\end{figure}

The plots in Fig.~\ref{fig:g11Results} are the same as those in
Fig.~\ref{fig:LambdaMassPlots}, except taken with the g11 dataset.
Here, the plots on the left are the invariant mass plots of the
$\pi^-p$ system, where only the invariant mass closest to
$m_{\Lambda}=1.115\,\mathrm{GeV}$ is plotted.  The scattered $\Lambda$
shows up very strongly as a peak above the background in the top left
plot.  For events with $m_{\Lambda_{s}}$ within 5~MeV of the known
value of $m_{\Lambda}=1.115\,\mathrm{GeV}$, the missing mass $m_X$ for
the process $Xp\to\Lambda p$ is shown in the right plots.  While the
background in the top right plot is admittedly very large, there is
clearly a peak above the background.  The dark line in the top plots
is a fit to a 3rd-order polynomial background.  In the bottom plots in
the Figure, the background has been subtracted, and the remaining peak
has been fit with a gaussian.  Note that the bottom right plot shows
17,000 events, which is more than an order of magnitude greater than
the existing world data sample.  The analysis of this dataset is still
underway.

\subsection{$pp$ Elastic Scattering}
For other tests of this technique, we chose to take advantage of the
skim performed in the context of the $\Lambda p$ rescattering pilot
study.  This skim required two protons in the final state.  With this
data skim, the simplest process that can be studied is $pp\to pp$.
This process, shown in Fig.~\ref{fig:PreviousDatapp},
has been measured very well; we do not expect to improve upon the
existing data set.  However, this is still a useful test of our cross
section determination for the $\Lambda p$ cross section.

\begin{figure}[h]
  \centerline{
    \includegraphics[width=0.45\textwidth]{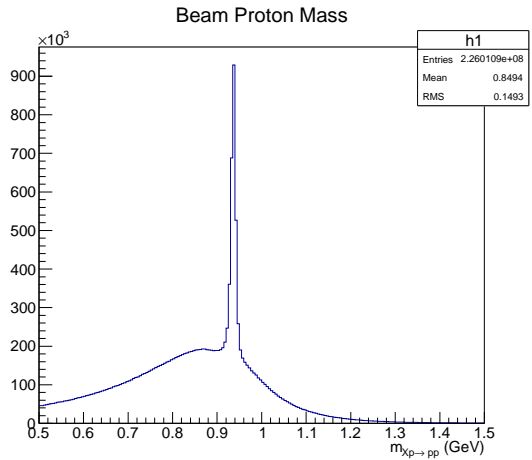}}
  \caption{\label{fig:ppMissingMass}The missing mass $m_X$ in the
    process $Xp\to pp$ for the g12 dataset.  The proton is strongly
    seen over a smooth background.}
\end{figure}

The analysis of $pp$ elastic scattering is slightly different than
that of $\Lambda p$ scattering, in that the scattered proton is
detected directly, and need not be reconstructed.  The only analysis
to be done here is to identify the two final-state protons, and to
look for a peak at the proton mass in the missing mass plot of the
process $Xp\to pp$.  This is shown in Fig.~\ref{fig:ppMissingMass}.

It should be pointed out here that the CLAS g12 trigger required three
charged particles.  Because of this, we do not have access to protons
produced in the process $\gamma p\to\pi^0p$.  Instead, the proton
production process is $\gamma p\to\rho^0p$ or $\gamma p\to\omega p$,
which clearly is still useful for studies of this sort.

\subsection{$K_sp$ Elastic Scattering}
Like the $\Lambda$, the $K_s$ is a very short-lived particle, with
$c\tau=2.68\,\mathrm{cm}$.\cite{RPP18}  There are even fewer cross
section measurements for the process $K_Sp\to K_Sp$ than for $\Lambda
p\to\Lambda p$; only one published measurement exists.\cite{And75}
The complete process in the CLAS detector is $\gamma p\to
K_S^0\Sigma^+; K_Sp\to K_Sp$.  The analysis of this process is
complicated by the fact that, since we do not detect the beam $K_s$,
there is contamination from the process $K_Lp\to K_Sp$, which will
need to be subtracted from our result.

\begin{figure}[h]
  \centerline{
    \includegraphics[width=0.45\textwidth]{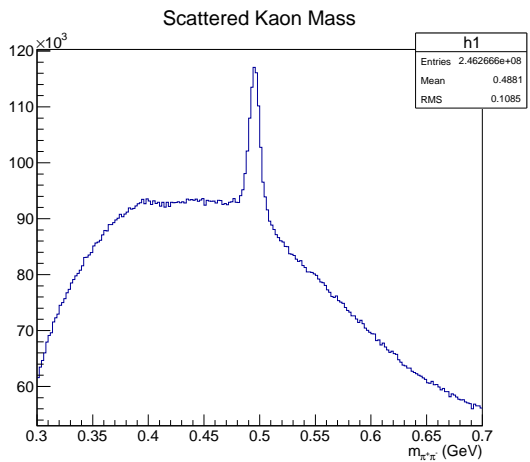}}
  \caption{\label{fig:KsInvariantMass}The invariant mass
    $m_{\pi^+\pi^-}$ of the $\pi^+\pi^-$ system in the g12 dataset.
    The $K_S$ is clearly seen over a smooth background.}
\end{figure}

We have begun the analysis of this process by looking for the $K_S$
via its decay to $\pi^+\pi^-$.  Figure~\ref{fig:KsInvariantMass} shows
the invariant mass spectrum of $\pi^+\pi^-$, showing a strong peak
above a smooth background.
This analysis is still in its infancy; we have yet to isolate the beam
$K_S$ in the process $K_Sp\to K_Sp$, which is the next step.

\subsection{$\Xi p$ Elastic Scattering}
As mentioned previously, this technique could be used to test the
Additive Quark Model of Levin and Frankfurt.\cite{Lev65}  At the
present time, we do not have a dataset large enough to see a useful
number of events for this process.  In the future, we hope to use the
upgraded CLAS12 detector to address this question.

\section{CONCLUSIONS}
We have shown that it is possible to create a beam of short-lived
particles to use in scattering processes on the proton.  We have used
this technique with the CLAS detector to produce a beam of $\Lambda$
hyperons, with which we have successfully identified the process
$\Lambda p\to\Lambda p$ in two independent datasets.  In so doing, we
have increased the world data sample for this process by well over an
order of magnitude.

Using the same dataset, we have shown that the process $pp\to pp$ is
also accessible via CLAS data, and can be used to check the systematic
uncertainty of the $\Lambda p\to\Lambda p$ cross section measurement.
We have also shown that the luminosity, while not yet completely
understood, is not an intractable problem.

Finally, this new technique opens up a whole new range of possible
studies with the CLAS detector that were not considered previously.
This will enable a large increase in the amount of physics output from
CLAS in the future.  There is a great deal of data available to test
out new methods, which can then be more finely tuned by the proposal
of dedicated experiments using this technique.

% Acknowledgement
\section{ACKNOWLEDGMENTS}
This work was performed with the support of a grant from the US
Department of Energy.  This program is being pursued in collaboration
with CSU Dominguez Hills undergraduate students Noraim Nu\~nez and
Marcos Guillen, and with Dr. Ken Hicks and Joey Rowley from Ohio
University. 

% References

\nocite{*}
\bibliographystyle{aipnum-cp}%
\bibliography{Price-HYP2018}%

\end{document}